\documentclass[conference]{IEEEtran}
\IEEEoverridecommandlockouts
\usepackage{cite}
\usepackage{enumitem}
\usepackage{amsmath,amssymb,amsfonts}
\usepackage{algorithmic}
\usepackage{graphicx}
\usepackage{textcomp}
\usepackage{xcolor}
\def\BibTeX{{\rm B\kern-.05em{\sc i\kern-.025em b}\kern-.08em
    T\kern-.1667em\lower.7ex\hbox{E}\kern-.125emX}}
    
\usepackage{acronym}
\newacro{IoT}[IoT]{internet-of-things}
\newacro{LEO}[LEO]{low-earth-orbit}
\newacro{LoS}[LoS]{line-of-sight}
\newacro{GS}[GS]{ground station}
\newacro{AF}[AF]{amplify-and-forward}
\newacro{SC}[SC]{selection combining}
\newacro{MRC}[MRC]{maximal ratio combining}
\newacro{SNR}[SNR]{signal-to-noise ratio}
\newacro{LPWA}[LPWA]{low-power wide-area}
\newacro{LoRa}[LoRa]{long range}
\newacro{NB-IoT}[NB-IoT]{narrowband-IoT}
\newacro{HSTN}[HSTN]{hybrid satellite terrestrial network}
\newacro{NOMA}[NOMA]{non-orthogonal multiple access}
\newacro{SS}[SS]{single satellite}
\newacro{AWGN}[AWGN]{additive white Gaussian noise}
\newacro{PDF}[PDF]{probability density function}
\newacro{CDF}[CDF]{cumulative distribution function}
\newacro{SR}[SR]{shadowed-Rician}
\newacro{EIRP}[EIRP]{equivalent isotropically radiated power}
\newacro{G/T}[G/T]{antenna gain-to-noise-temperature}
\newacro{OP}[OP]{outage probability}
\newacro{DF}[DF]{decode-and-forward}
\begin{document}

\title{Performance Analysis of Novel Direct Access Schemes for LEO Satellites Based IoT Network \\
}

\author{Ayush Kumar Dwivedi\IEEEauthorrefmark{1}, Sai Praneeth Chokkarapu\IEEEauthorrefmark{1}, Sachin Chaudhari\IEEEauthorrefmark{1}, Neeraj Varshney\IEEEauthorrefmark{2}\\
{\textit{\IEEEauthorrefmark{1}International Institute of Information Technology Hyderabad, 500032 India}} \\
\textit{\IEEEauthorrefmark{2}Wireless Networks Division, National Institute of Standards and Technology, Gaithersburg, MD 20899 USA}\\
ayush.dwivedi@research.iiit.ac.in, saipraneeth.c@students.iiit.ac.in, sachin.c@iiit.ac.in, neerajv@ieee.org}

\maketitle

\begin{abstract}


This paper analyzes the performance of low earth orbit (LEO) satellites based internet-of-things (IoT) network where each IoT node makes use of multiple satellites to communicate with the  ground station (GS). In this work, we consider fixed and variable gain amplify-and-forward (AF) relaying protocol at each satellite where the received signal from each IoT node is amplified before transmitting to the terrestrial GS for data processing. To analyze the performance of this novel LEO satellites based direct access architecture, the closed-form expressions for outage probability are derived considering two combining schemes at the GS: \textit{(i)} selection combining; \textit{(ii)} maximal ratio combining. Further, to gain more insights for diversity order and coding gain, asymptotic outage probability analysis at high SNR for both schemes is also performed. Finally, simulation results are presented to validate the analytical results derived and also to develop several interesting insights into the system performance. 
\end{abstract}
\vspace{-0.3cm}
\begin{IEEEkeywords}
Satellite based IoT, LEO satellites, Amplify-and-Forward, outage probability
\end{IEEEkeywords}

\vspace{-0.3cm}
\section{Introduction}
\label{intro}
With a plan to launch 60 satellites every two weeks at 10 times lesser cost, SpaceX Starlink has proved that launching \ac{LEO} satellites is no more a rocket science~\cite{spacex}. The LEO constellations like Starlink, OneWeb, Iridium, Telesat and many more under development have started a new era of affordable satellite communication. Out of the expected 20 billion connected \textit{things} by the end of 2020, an estimated 5.3 million connections will be through satellite services~\cite{iot}. Although \ac{LPWA} networks like \ac{LoRa} and \ac{NB-IoT} are designed to cater to the low-data low-power requirements of \ac{IoT}, these technologies fail to provide global coverage and are susceptible to natural calamities. For example, 4G wireless network currently covers 63\% of world population (same expected for 5G network) but only 37\% of landmass~\cite{nicolas}. On contrary, satellite-based access network can provide global coverage to \ac{IoT} devices which are often deployed at remote locations and are dispersed over a large geographical area. In particular, large-scale \ac{LEO} satellites have proved their potential in addressing coverage issues~\cite{direct, survey}. For example, \ac{IoT} for smart operations such as farms, oil/gas installations, electric grid etc. can benefit from satellite by extending the terrestrial coverage~\cite{gopal}.

Integration of satellites for use in \ac{IoT} networks is done in two modes~\cite{modes}: \textit{direct} access and \textit{indirect} access. In direct access mode, the \ac{IoT} devices communicate with the satellite directly, while in indirect mode, the IoT devices connect with satellite through a terrestrial \ac{LPWA} gateway/relay. Such gateways have small aperture satellite terminals as well as traditional terrestrial \ac{LPWA} radio modules. However, the use of indirect access is limited by the coverage of the terrestrial gateway. Investing in gateways is also not profitable in applications deployed at locations hit by disaster or locations requiring deployments for short duration. In contrast to this, direct access mode is an appealing solution for such scenarios. Thanks to the communication modules developed by companies like Iridium, Kepler and Hiber, many low powered radio modules are available in commercial market which offer direct access satellite communication from \ac{IoT} nodes.

Another critical aspect of \ac{IoT} applications is that, these are sensing heavy. Offloading the sensed information to a data centre is a major task. For example, use-cases like air pollution monitoring or intrusion detection involves sending status data at regular or random intervals of time. \ac{IoT} devices in such scenarios do not have much computational resources to execute complex gateway selection or scheduling algorithms. For such applications, \ac{LoRa} WAN technologies with star-of-stars topology became famous in short time~\cite{lora}. They makes use of a gateways which act like transparent bridges between the end-devices and the central data server. The \ac{IoT} nodes broadcast their sensed information which is typically heard by multiple gateways. The central server selects information from one of the gateways and ignore the others. In a nutshell, using such an architecture, even the dumb nodes can communicate the sensed information to the data server. It is important to devise similar architectures which suit \ac{IoT} requirements, but at the same time provides global coverage as well. Satellites come to rescue in this case yet again. By virtue of orbital dynamics, they can naturally support broadcast, multicast or geocast transmissions~\cite{direct}, where each satellite can act as a transparent bridge between the node and the central data server. This paper envisions a new architecture for direct access based \ac{LEO} satellite IoT network inspired from the topology used in terrestrial \ac{LoRa} networks. Constellations of \ac{LEO} satellites acting as transparent relays/bridges prove to be an invaluable resource in this architecture. 

In the past, many studies have been done to evaluate the performance of direct and indirect access based satellite systems. A comprehensive survey on use of satellites for \ac{IoT} network is presented in \cite{survey}. The work in \cite{sreng} studied a relay cooperation based \ac{HSTN}. Authors evaluated a \ac{NOMA} assisted overlay \ac{HSTN} in \cite{overlay}. Performance of a dual-hop satellite relaying system for multiple users is analysed in \cite{dualhop}. 

This is an introductory work and intends to establish the motivation for the idea discussed. The main contribution of this paper is described below:
\begin{itemize}
    \item An architecture is proposed which suits sensing heavy \ac{IoT} applications and also harvests the benefits of existing and upcoming mega \ac{LEO} constellations. In this architecture, fixed and variable gain \ac{AF} relaying protocol is employed at each satellite where the received signal from each \ac{IoT} node is amplified before transmission to the terrestrial \ac{GS} for data processing.
    \item Performance of the proposed architecture in terms of \ac{OP} is quantified. For this purposes, we derive closed-form analytical expressions for two different combining schemes i.e., \ac{SC} and \ac{MRC} at \ac{GS}.
    \item Asymptotic analysis of \ac{OP} at high \ac{SNR} to gain more insights into the diversity order and coding gain is done.
\end{itemize}
The rest of the paper is organised as following. The complete system model is described in Section \ref{sys}. This is followed by the exact and asymptotic outage probability analysis in Section \ref{op_anls}. Simulation results are presented in Section \ref{sim}, followed by the conclusion in Section \ref{sec:conc}.

\section{System Description}
\label{sys}
\begin{figure}[t!]
\centerline{\includegraphics[width=3.0in, height=2.5in]{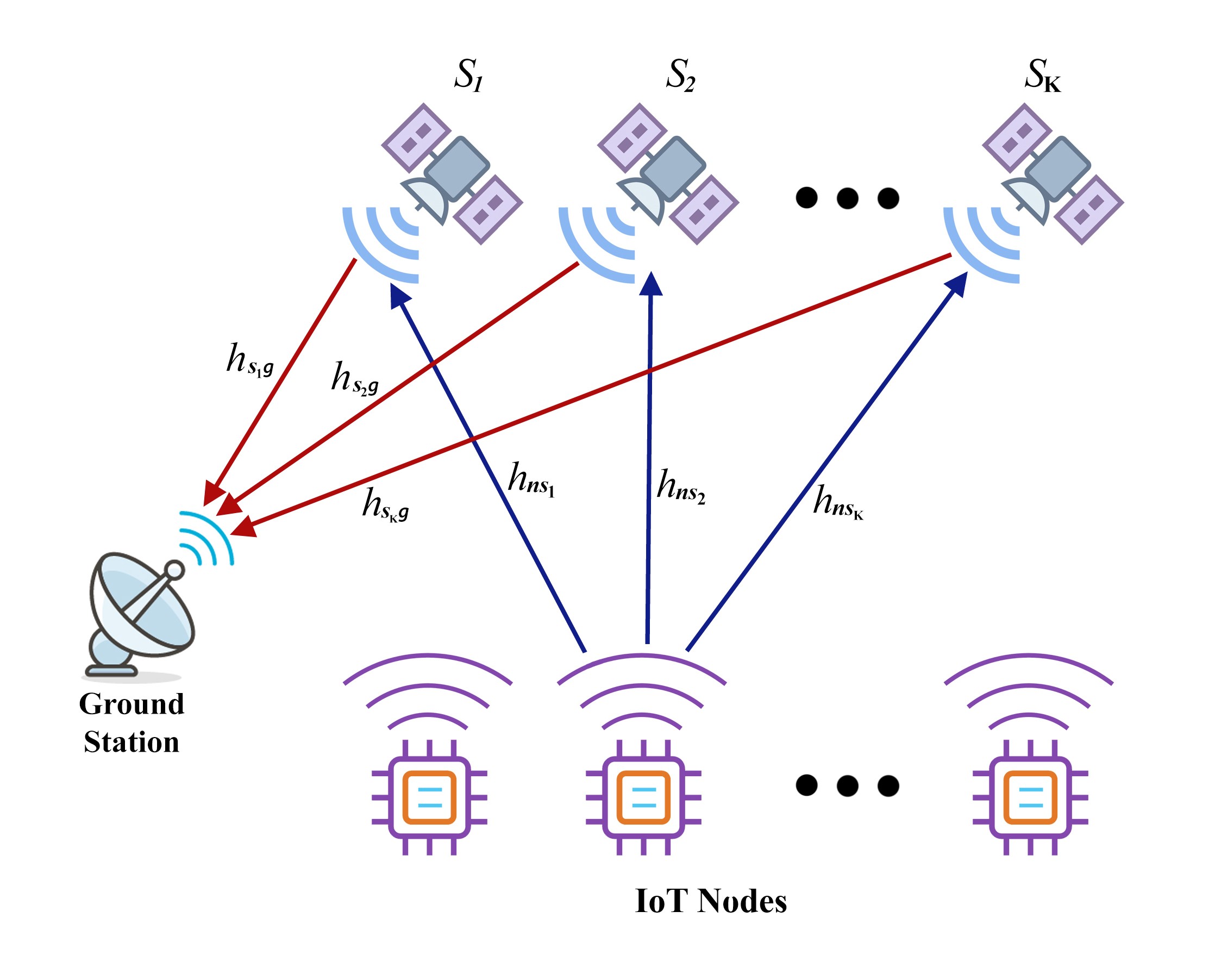}}
\vspace{-0.6cm}
\caption{Schematic diagram of the proposed \ac{LEO} satellite based direct access network where multiple \ac{IoT} nodes which are in \ac{LoS}, communicate to \ac{GS} using $K$ \ac{LEO} satellites.}
\vspace{-0.5cm}
\label{sm}
\end{figure}

A \ac{LEO} satellite based direct access architecture is considered where a set of terrestrial \ac{IoT} nodes communicate with a \ac{GS} using $K$ \ac{LEO} satellites ${\{S_k\}}_{k=1}^{K}$, as shown in Fig. \ref{sm}.  Similar to the \ac{NB-IoT} where dedicated sub-carriers of $3.75$ kHz or $15$ kHz are utilised, it is assumed that these \ac{IoT} nodes utilize separate frequency carriers such that there is no interference from adjacent node \cite{link2, protocols}. Hence analysis is performed for a single \ac{IoT} node and the same can be extended for other nodes as well. As per the proposed system model, the \ac{IoT} node broadcasts its information to all the $K$ satellites. The signal received at each of the $K$ \ac{LEO} satellites is amplified and forwarded to a terrestrial \ac{GS} for data-processing.

We compare the performance of this architecture in terms of outage probability at the \ac{GS} for three different schemes.
\begin{itemize}
    \item Scheme-1: Information is decoded at \ac{GS} using the signal received from a \ac{SS} only.
    \item Scheme-2: Information is decoded at \ac{GS} using \ac{SC} scheme, where a strong signal out of $K$ received signals from $K$ \ac{LEO} satellites is selected for decoding. 
    \item Scheme-3: Information is decoded at \ac{GS} after coherently combining of $K$ received signals using \ac{MRC} technique.
\end{itemize}

The end-to-end communication between the \ac{IoT} node and the \ac{GS} takes place in two phases. In the first phase, the \ac{IoT} node broadcasts its information signal to $K$ satellites. The received signal at the $k$th satellite can be written as
\begin{equation}
y_{ns_k} = \sqrt{P_{n}} h_{ns_k} x_n + n_{ns_k},
\label{y1}
\end{equation}
where $P_{n}$ is the transit power of the IoT node, $h_{ns_k}$ is the coefficient of the channel between the IoT node and the $k$th satellite, $x_n$ is the unit energy information signal and $n_{ns_k}$ the additive noise modeled as independent and identically distributed (i.i.d.) symmetric complex Gaussian with mean zero and variance $\sigma^2$.  

In the second phase, each satellite relays the signal received from \ac{IoT} node to \ac{GS} by employing \ac{AF} relaying scheme. Therefore, the signal received at \ac{GS} from $k$th satellite can be written as
\begin{equation}
y_{s_kg} = \sqrt{P_{s_k}} h_{s_kg} \mathbb{G} (\sqrt{P_{n}} h_{ns_k} x_n + n_{ns_k}) + n_{s_kg},
\label{y2}
\end{equation}
where $P_{s_k}$ is the transit power of the $k$th satellite, $h_{s_kg}$ is the coefficient of the channel between $k$th satellite and \ac{GS}, $\mathbb{G}$ is the \ac{AF} gain factor and $n_{s_kg}$ is the \ac{AWGN} noise at \ac{GS} receiver.  Note that all the channel coefficients corresponding to node-satellite and satellite-GS links i.e., $h_{ns_k}$  and $h_{s_kg}, \forall k,$ are assumed to follow \ac{SR} distribution. The \ac{SR} fading model is best known for characterizing communication links which suffer from \ac{LoS} shadowing and small scale fading. It is a more generalized form of Rician fading model where the amplitude of the \ac{LoS} component follows Nakagami-$m$ fading. Moreover, this model is widely accepted for characterizing satellite channels and fits the experimental data very well \cite{sr}.

Using \eqref{y1}, the instantaneous \ac{SNR} at the $k$th satellite for the node-satellite link is given by $\Lambda_{ns_k} = P_n|h_{ns_k}|^2/\sigma ^2 = \eta_{n} |h_{ns_k}|^2,$
where $\eta_{n} = \frac{P_n}{\sigma^2}$. Further, under scheme-1 (\ac{SS}), the instantaneous \ac{SNR} at \ac{GS} corresponding to the transmission by $k$th satellite is given by
\begin{equation}
\Lambda_{\text{GS}}^{\text{SS}} = \frac{\Lambda_{s_kg}\ \Lambda_{ns_k}}{\Lambda_{s_kg} + C},
\label{snr_sch1}
\end{equation}
where $\Lambda_{s_kg} = \eta_{s_k} |h_{s_kg}|^2$, $\eta_{s_k} = P_{s_k}/\sigma^2$ and $C = 1/\mathbb{G}^2 \sigma^2$. The \ac{AF} gain factor $\mathbb{G}$ is defined as \cite{max_max}
\begin{equation}
\vspace{-0.1cm}
\mathbb{G} = \left( P_n|h_{ns_k}|^2 + \sigma^2\right)^{-\frac{1}{2}}.
\label{gain}
\end{equation}
In \eqref{snr_sch1}, the term $C$ can be simplified to $1 + \Lambda_{ns_k}$ or $1 + \mathbb{E}[\Lambda_{ns_k}]$ depending upon the choice of variable gain or fixed gain relaying~\cite{max_max}. Similarly, the instantaneous \acp{SNR} at \ac{GS} under scheme-2 (\ac{SC}) and scheme-3 (\ac{MRC}) are given, respectively,~by
\vspace{-0.1cm}
\begin{align}
    \Lambda_{\text{GS}}^{\text{SC}} =& \max_{k}\left(\frac{\Lambda_{s_kg}\ \Lambda_{ns_k}}{\Lambda_{s_kg} + C}\right),
\label{snr_sch2} \\
\Lambda_{\text{GS}}^{\text{MRC}} =& \frac{\sum_{k=1}^{K}(\Lambda_{s_kg})\ \sum_{k=1}^{K} (\Lambda_{ns_k})}{\sum_{k=1}^{K}(\Lambda_{s_kg}) + C_m},
\label{snr_sch3}
\end{align}
where $C_m$ is defined as $C_m = 1/(\sigma^2\sum_{k=1}^{K}\mathbb{G}^2)$.

\subsection{Statistical characteristics of shadowed-Rician channel}\label{3A}
The \ac{PDF} and \ac{CDF} of $\Lambda_{i} = \eta_i |h_{i}|^2$, $i \in \{ns_k, s_kg\}$ and $k \in \{1,2,\ldots,K\}$ are given, respectively, by \cite{overlay}
\vspace{-0.1cm}
\begin{align}
    f_{\Lambda_{i}(x)} = &\alpha_i \sum_{\kappa = 0}^{m_i-1} \frac{\zeta(\kappa)}{\eta_i^{\kappa + 1}} x^{\kappa} e^{-\left(\frac{\beta_i - \delta_i}{\eta_i} \right)x},
\label{pdf_sr} \\
F_{\Lambda_{i}(x)} =& 1 - \alpha_i \sum_{\kappa = 0}^{m_i-1} \frac{\zeta(\kappa)}{\eta_i^{\kappa + 1}} \sum_{p=0}^{\kappa}\frac{\kappa!}{p!} \left(\frac{\beta_i - \delta_i}{\eta_i} \right)^{-(\kappa + 1 - p)}\nonumber \\
& \times x^p  e^{-\left(\frac{\beta_i - \delta_i}{\eta_i} \right)x},
\label{cdf_sr}
\end{align}
where $\alpha_i = ((2b_i m_i)/(2b_i m_i +\Omega_i))^{m_i}/2b_i$, $\beta_i = 1/2b_i$, $\delta_i = \Omega_i/(2b_i)(2b_i m_i + \Omega_i)$ and $\zeta(\kappa) = (-1)^{\kappa} (1-m_i)_{\kappa} \delta_i^{\kappa}/(\kappa!)^2$ with $(\cdot)_\kappa$ being the Pochhammer symbol\cite{formula}. Here $2b_i$ denotes the average power of the multipath component, $\Omega_i$ is the average power of \ac{LoS} component
\vspace{-0.1cm}
\section{Outage Performance of \ac{LEO} satellite based direct access \ac{IoT} network}
\vspace{-0.1cm}
\label{op_anls}
In this section the performance of the proposed architecture is analyzed. For this purpose, closed-form expressions for outage probability in all the three schemes are derived.
\vspace{-0.1cm}

\subsection{Outage Probability Analysis}

\textit{1) Scheme-1 (\ac{SS}):} The outage probability at \ac{GS} in the case of single satellite can be evaluated as
\begin{equation}
\begin{split}
P_{\text{out}}^{\text{SS}}(R) & = \Pr\left[\frac{1}{2}\log_2(1+\Lambda_{\text{GS}}^{\text{SS}}) \leq R\right]  \\
 & = \Pr\left[\frac{\Lambda_{sg}\ \Lambda_{ns}}{\Lambda_{sg} + C} \leq \gamma_{\text{th}}\right],
\end{split}
\label{op_ss}
\end{equation}
where $R$ is the target rate and $\gamma_{\text{th}} \triangleq  2^{2R} - 1$. We can reformulate \eqref{op_ss} under variable gain relaying as
\begin{equation}
P_{\text{out}}^{\text{SS}}(R) = \Pr\left[ (\Lambda_{sg} - \gamma_{\text{th}}) (\Lambda_{ns} - \gamma_{\text{th}}) \leq \gamma_{\text{th}}^2 + \gamma_{\text{th}}\right].
\label{op_ss1}
\end{equation}
Further, using \eqref{pdf_sr}, the above expression is mathematically intractable and difficult to solve in closed-form. Hence, we employ an $M$-step-staircase approximation approach as in \cite{step}. Using this approach, the closed-form expression for $P_{\text{out}}^{\text{SS}}(R)$ is derived as \eqref{op_ss_closed}, where $\Upsilon = \gamma_{\text{th}}^2 + \gamma_{\text{th}}$. The detailed proof of \eqref{op_ss_closed} is given in Appendix \ref{AppA}.

\begin{figure*}[!t]
\normalsize
\newcounter{mycnt1}
\setcounter{mycnt1}{\value{equation}}
\setcounter{equation}{10}
\begin{align}
\label{op_ss_closed}
P_{\text{out}}^{\text{SS}}&(R)  = F_{\Lambda_{sg}}(\gamma_{\text{th}}) + 
\left\{F_{\Lambda_{ns}}(\gamma_{\text{th}})\times\left[1 -   F_{\Lambda_{sg}}(\gamma_{\text{th}}) \right]\right\} + \left\{\left[F_{\Lambda_{sg}}(\sqrt{\Upsilon} + \gamma_{\text{th}}) - F_{\Lambda_{sg}}(\gamma_{\text{th}})\right] {\times} \left[F_{\Lambda_{ns}}(\sqrt{\Upsilon} + \gamma_{\text{th}}) - F_{\Lambda_{ns}}(\gamma_{\text{th}})\right]\right\}\nonumber\\
& + \sum_{i=1}^{M}\left\{ \left[F_{\Lambda_{sg}}\left(\frac{\Upsilon}{\sqrt{\Upsilon} + \frac{i-1}{M}L} + \gamma_{\text{th}}\right) - F_{\Lambda_{sg}}(\gamma_{\text{th}})\right] {\times} \left[F_{\Lambda_{ns}}\left(\sqrt{\Upsilon} + \gamma_{\text{th}} + \frac{iL}{M}\right) - F_{\Lambda_{ns}}\left(\sqrt{\Upsilon} + \gamma_{\text{th}} + \frac{(i-1)L}{M}\right)\right]\right\}\nonumber\\
& + \sum_{i=1}^{M}
\left\{ \left[F_{\Lambda_{ns}}\left(\frac{\Upsilon}{\sqrt{\Upsilon} {+} \frac{i-1}{M}L} {+} \gamma_{\text{th}}\right) {-} F_{\Lambda_{ns}}(\gamma_{\text{th}})\right] {\times} \left[F_{\Lambda_{sg}}\left(\sqrt{\Upsilon} {+} \gamma_{\text{th}} + \frac{iL}{M}\right) {-} F_{\Lambda_{sg}}\left(\sqrt{\Upsilon} {+} \gamma_{\text{th}} {+} \frac{(i{-}1)L}{M}\right)\right]\right\},\!\!
\end{align}
\vspace{-0.1cm}
\hrule
\vspace{-0.1cm}
\setcounter{equation}{16}
\begin{align}
\label{op_mrc_closed}
\!\!P_{\text{out}}^{\text{MRC}}&(R)  = F_{\Delta_{ns}}(\gamma_{\text{th}}) + \left\{F_{\Delta_{sg}}(\sqrt{C_{m} \gamma_{\text{th}}}) \times \left[F_{\Delta_{ns}}\left(\sqrt{C_{m}\gamma_{\text{th}}} + \gamma_{\text{th}}\right) - F_{\Delta_{ns}}\left(\gamma_{\text{th}}\right)\right]\right\}\nonumber\\
& + \sum_{i=1}^{M}\left\{F_{\Delta_{sg}}\left(\frac{C_{m}\gamma_{\text{th}}}{\sqrt{C_{m}\gamma_{\text{th}}} + \frac{i-1}{M}L}\right) {\times}
\left[F_{\Delta_{ns}}\left(\sqrt{C_{m}\gamma_{\text{th}}} + \gamma_{\text{th}} + \frac{i}{M}L\right) - F_{\Delta_{ns}}\left(\sqrt{C_{m}\gamma_{\text{th}}} + \gamma_{\text{th}} + \frac{i-1}{M}L\right)\right]\right\}\nonumber\\
& + \sum_{i=1}^{M}\left\{\left[ F_{\Delta_{sg}}\left(\sqrt{C_{m}\gamma_{\text{th}}} {+} \frac{i}{M}L\right) {-} F_{\Delta_{sg}}\left(\sqrt{C_{m}\gamma_{\text{th}}} {+} \frac{i{-}1}{M}L\right)\right]{\times} 
\left[F_{\Delta_{ns}}\left(\frac{C_{m}\gamma_{\text{th}}}{\sqrt{C_{m}\gamma_{\text{th}}} {+} \frac{i{-}1}{M}L} {+} \gamma_{\text{th}}\!\right) {-} F_{\Delta_{ns}}(\gamma_{\text{th}}) \right]\!\right\},\!\!
\end{align}
\stepcounter{mycnt1}
\setcounter{equation}{\value{mycnt1}}
\vspace{-0.1cm}
\hrule
\vspace{-0.6cm}
\end{figure*}

\textit{2) Scheme-2 (\ac{SC}):}
The outage probability at \ac{GS} in the case of \ac{SC} can be evaluated as
\begin{equation}
P_{\text{out}}^{\text{SC}}(R) =\Pr\left[\max_{k}\left(\frac{\Lambda_{s_kg}\ \Lambda_{ns_k}}{\Lambda_{s_kg} + C}\right) \leq \gamma_{\text{th}}\right].
\label{op_sc}
\end{equation}
Further, following the similar analysis as done in the case of scheme-1, the closed form expression for \ac{SC} under variable gain \ac{AF} relaying can be derived as
\begin{equation}
P_{\text{out}}^{\text{SC}} = \prod_{k=1}^{K} P_{\text{out,k}}^{\text{SS}}(R),
\label{op_sc_final}
\end{equation}
where $P_{\text{out,k}}^{\text{SS}}(R)$ is given in \eqref{op_ss_closed}.

\textit{3) Scheme-3 (MRC):} Using \eqref{snr_sch3}, the outage probability in the case of MRC is given by
\begin{equation}
P_{\text{out}}^{\text{MRC}}(R) = \Pr\left[\frac{\Delta_{sg}\ \Delta_{ns}}{\Delta_{sg} + C_m}\leq \gamma_{\text{th}}\right],
\label{op_mrc}
\end{equation}
where $\Delta_{sg}$ and $\Delta_{ns}$ are defined as $\Delta_{sg} \triangleq \sum_{k=1}^{K}(\Lambda_{s_kg})$ and $\Delta_{ns} \triangleq \sum_{k=1}^{K}(\Lambda_{ns_k})$. The CDF of $\Delta_{i}, i\in\{sg,ns\}$ is given in Appendix \ref{AppB}. Due to analytical tractability, we reformulate \eqref{op_mrc} under fixed gain relaying as
\begin{equation}
P_{\text{out}}^{\text{MRC}}(R) = \Pr\left[ (\Delta_{sg}) (\Delta_{ns} - \gamma_{\text{th}}) \leq C_m\gamma_{\text{th}}\right],
\label{op_mrc1}
\end{equation}
where $C_m$ is given as $\left[\sum_{k=1}^{K}1/(1 + \mathbb{E}\left[\Lambda_{ns_k}\right])\right]^{-1}$. It is worth-mentioning that $\mathbb{E}\left[\Lambda_{ns_k}\right]$ can be derived analytically by calculating the expectation over \eqref{pdf_sr}. The derived simplified expression for $\mathbb{E}\left[\Lambda_{i}\right]$ is given by
\begin{equation}
\mathbb{E}\left[\Lambda_i\right] = \alpha_i \sum_{\kappa = 0}^{m_i-1}\zeta(\kappa)\eta_i \frac{\Gamma(\kappa+2)}{(\beta_i - \delta_i)^{\kappa + 2}}.
\label{e_sr}
\end{equation}
Further, after applying the $M$-step-staircase approximation for \eqref{op_mrc1}, the closed-form expression for outage probability in case of \ac{MRC} can be derived as \eqref{op_mrc_closed}. The proof can be carried out by following the similar steps as shown in Appendix \ref{AppA}.

\vspace{-0.1cm}
\subsection{Asymptotic Outage Probability Analysis}
\vspace{-0.1cm}
In this section, we provide the asymptotic outage probability analysis under high \ac{SNR} assumption. This will help us to gain more insight about the network in terms of diversity order that it can achieve.

At high \ac{SNR} i.e., $\eta_n, \eta_{s_k} \rightarrow \infty$, the \acp{CDF} used in \eqref{op_ss_closed} and \eqref{op_mrc_closed} can be approximated respectively as in \cite{dualhop} by
\setcounter{equation}{17}
\begin{align}
F_{\Lambda_i}^{\infty}(x) \approx &\frac{\alpha_i}{\eta_i}x, ~~\text{and}~~~ F_{\Delta_i}^{\infty}(x) \approx& \frac{\alpha_i^K}{\eta_i^K \Gamma(K+1)}x^K.
\label{cdf_ss_asymp}
\end{align}
The total transmit power of the system, $P_t$ can be written as $P_t = P_n + \sum_{k=1}^{K}P_{s_k}$. Considering equal power allocation, the asymptotic outage probability for scheme-2 and scheme-3 can be derived respectively,~as 
\begin{align}
P_{\text{out}}^{\text{SC},\infty} =&\left( \frac{\gamma_{\text{th}} }{\eta}\prod_{k=1}^{K}\sqrt[K]{\left(\alpha_{s_kg}{+}\alpha_{ns_k}\right)}\right)^K {+} O\left(\frac{1}{\eta^K}\right),
\label{sc_assymp}\\
P_{\text{out}}^{\text{MRC},\infty} =&\left( \frac{\gamma_{\text{th}} \alpha_{ns}}{\sqrt[K]{\Gamma(K+1)}\eta}\right)^K + O\left(\frac{1}{\eta^K}\right),
\label{mrc_assymp}    
\end{align}
where $O(\cdot)$ stands for higher order terms. The outage probability at high \ac{SNR} can be approximated as
\begin{equation}
P_{\text{out}}^\infty = (G_c\eta)^{-d} + O(\eta^{-d}),
\label{op_assymp}
\end{equation}
where $G_c$ and $d$ denote the coding gain and the diversity order, respectively. Hence, by comparing \eqref{sc_assymp}, \eqref{mrc_assymp} and \eqref{op_assymp}, the diversity order of the system for both scheme-2 and scheme-3 can be seen as $K$. Moreover, the respective coding gains are
\begin{align}
G_c^{\text{SC}} = &\bigg(\gamma_{\text{th}} \prod_{k=1}^{K}\sqrt[K]{\left(\alpha_{s_kg}+\alpha_{ns_k}\right)}\bigg)^{-1}, \\ G_c^{\text{MRC}} = &\sqrt[K]{\Gamma(K+1)} (\gamma_{\text{th}} \alpha_{ns})^{-1}. 
\label{cg}
\end{align}

\section{Simulation Results}
\label{sim}
This section presents simulation results to validate the derived analytical results of this work and to develop several important insights into the system performance. For simulation purpose, we consider the following four possible shadowing conditions for all the three schemes:
\begin{enumerate}[label=(\alph*)]
\item H-H: each $h_{ns_k}$ and $h_{s_kg}$ under heavy shadowing (H)
\item H-A:  each $h_{ns_k}$ under heavy shadowing (H) and each $h_{s_kg}$ under average shadowing (A)
\item A-H: each $h_{ns_k}$ under average shadowing (A), each $h_{s_kg}$ under heavy shadowing (H)
\item A-A: each $h_{ns_k}$ and $h_{s_kg}$ under average shadowing (A)
\end{enumerate}
The \ac{SR} fading parameters $(m, b, \Omega)$ under heavy and average shadowing conditions are considered to be ($2, 0.063, 0.0005$) and ($5, 0.251, 0.279$), respectively~\cite{ch_value}. We set the target rate $R = 0.5$ so that $\gamma_{\text{th}} = 1$ \cite{overlay}. For step-staircase approximation, we set $M = 50$ and $L = 15\gamma_{\text{th}}$. For simplicity in simulation, we consider equal power allocation with $\eta_n = \eta_{s_k} = \eta$ as the transmit \ac{SNR}.

Further, to be able to identify the range of \ac{SNR} values which are feasible for proposed \ac{IoT} network, we performed the link budget analysis as given in \cite{link1}. For link budget, we consider a \ac{LEO} satellite at an altitude of 800 km, uplink central frequency of 950 MHz, satellite elevation angle of $30^\circ$, 3GPP Class 3 \ac{IoT} node transmit \ac{EIRP} of 23 dBm \cite{3gpp_class3}, and sub-carrier bandwidths of 3.7 kHz , 15 kHz, 45 kHz, 90 kHz and 180 kHz. For \ac{LEO} satellite receiver \ac{G/T} varying from -25 dBi/K to -6 dBi/K, \ac{SNR} in the approximate range of -9 dB to 20 dB are found feasible for our network. Thus, this \ac{SNR} range is considered for simulations under all the four possible channel conditions. 
\begin{figure}[t!]
\centerline{\includegraphics[width=3.2in,height=2.3in]{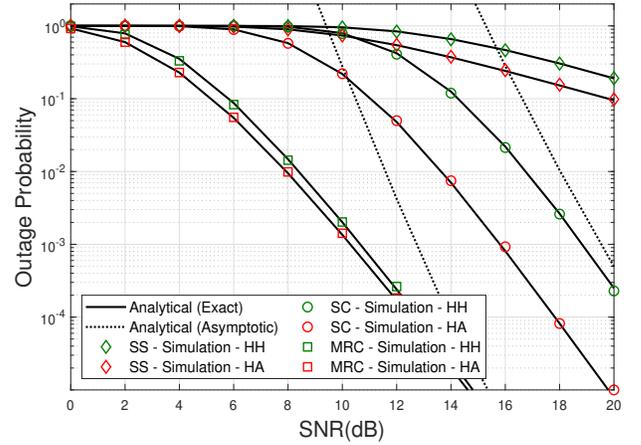}}
\vspace{-0.3cm}
\caption{\ac{OP} versus \ac{SNR} curves for all three schemes under H-H and H-A conditions using $\gamma_{\text{th}} = 1, M=50, L= 15\times\gamma_{\text{th}}$, and $K=5$.}
\vspace{-0.7cm}
\label{result1}
\end{figure}

Fig. \ref{result1} shows the outage performance of the system against \ac{SNR} by considering five \ac{LEO} satellites in \ac{LoS}. This figure considers shadowing conditions (a) and (b). We are using \ac{AF} relaying where the signal is amplified without decoding as opposed to \ac{DF} where the effect of shadowing can be un-done by decoding at the relay. As a result, the effect of node-satellite link shadowing is clearly visible on the performance. In this figure, node-satellite link is under heavy shadowing, consequently the outage probability is significantly high for \ac{SNR} less than 0 dB. Hence the outage probability curves are plotted for \ac{SNR} ranging from 0 dB to 20 dB in this figure to validate our closed form expressions. It can be seen that the \ac{MRC} and \ac{SC} schemes outperforms the \ac{SS} scheme. This proves the superiority of the proposed architecture and establishes how it is able to harvest the benefits of multiple satellites available as part of mega \ac{LEO} constellations. One can also observe that \ac{MRC} performs significantly better than \ac{SC}. Approximately $6$dB higher \ac{SNR} is required in case of \ac{SC} when compared to \ac{MRC}, for an outage probability of $10^{-2}$. Another key observation is that, in the case of \ac{SC}, approximately $3$dB higher \ac{SNR} is required when the satellite-\ac{GS} channel changes from average to heavy. Whereas such drastic effect of change in channel shadowing on outage performance is not seen in the case of \ac{MRC}. Therefore, \ac{MRC} proves to be much more robust towards change in shadowing conditions of satellite-\ac{GS} channel. The asymptotic curves approach the exact analytical curves sharply, thus validating the correctness of the derived formulae. It is also observed that the slope of \ac{OP} curves for \ac{SC} and \ac{MRC} are similar towards higher \ac{SNR}. This is indicative of the fact that diversity order depends on the number of satellites in \ac{LoS}.
\begin{figure}[t!]
\centerline{\includegraphics[width=3.2in,height=2.3in]{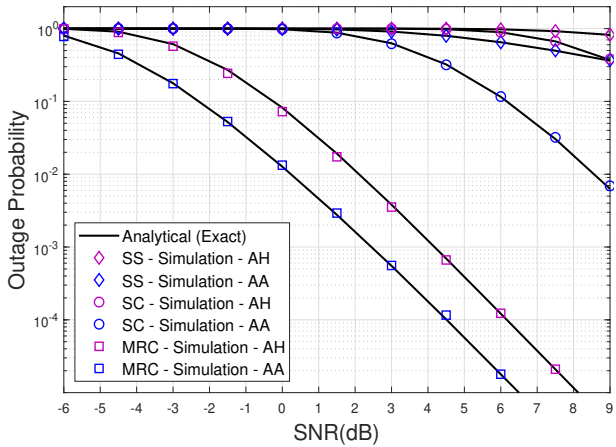}}
\vspace{-0.3cm}
\caption{\ac{OP} versus SNR curves for all three schemes under A-H and A-A conditions using $\gamma_{\text{th}} = 1, M=50, L= 15\times\gamma_{\text{th}}$, and $K=5$.}
\vspace{-0.6cm}
\label{result2}
\end{figure}

Fig. \ref{result2} shows the outage performance for shadowing conditions (c) and (d). Here the node-satellite channel is considered to be under average shadowing. Consequently as the result of \ac{AF} relaying, the \ac{OP} is significantly low for \ac{SNR} greater than 9 dB. Hence \ac{OP} curves are plotted for \ac{SNR} ranging from -6 dB to 9 dB in this figure to validate our closed form expressions. Observations similar to Fig. \ref{result1} can be made regarding the better performance of \ac{MRC} when compared to \ac{SC}. Although \ac{MRC} still proves to be more robust than \ac{SC} towards change in satellite-\ac{GS} link shadowing conditions, the difference in \ac{SNR} is high compared to Fig. \ref{result1}.

\begin{figure}[t!]
\centerline{\includegraphics[width=3.2in,height=2.3in]{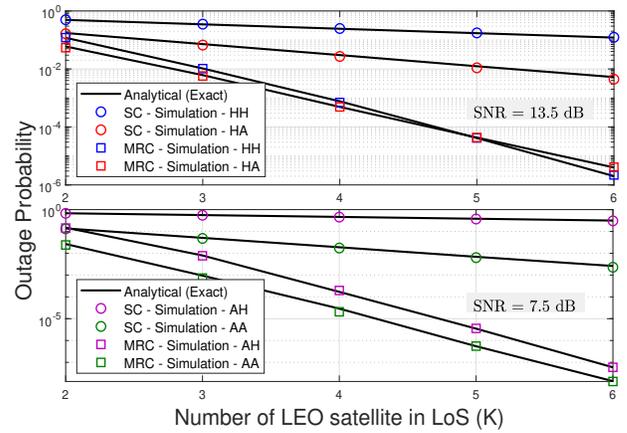}}
\vspace{-0.3cm}
\caption{\ac{OP} versus number of satellites in LoS (K) for scheme-2 and scheme-3 using $\gamma_{\text{th}} = 1, M=50, L= 15\times\gamma_{\text{th}}$ and $\eta = 13.5$dB for H-A, H-H conditions and $\eta = 7.5$dB for A-H, A-A conditions.}
\vspace{-0.45cm}
\label{result3}
\end{figure}

To gain more insights about the architecture, we also plot the outage probability versus number of satellites ($K$) in Fig.\ref{result3}. A comparison between three famous commercial LEO constellations (SpaceX-Starlink, OneWeb, Telesat) in terms of possible number of satellites in \ac{LoS} is given in \cite{consti}. According to that, for latitudes where majority of world's population is located, 2 to 30 LEO satellites can be in \ac{LoS} based on location of user equipment. Hence we simulate for $K$ in the range of 2 to 6. It can be seen from the plot that the system performance can be significantly enhanced by increasing $K$. This means that the performance of proposed architecture would increase as and when new constellations or satellites are added in network. It is also observed that system utilizing \ac{MRC} benefits more when compared to \ac{SC} with increase in number of satellites under all channel conditions.

\section{Conclusion}\label{sec:conc}
With the advent of mega \ac{LEO} constellations and direct access \ac{LEO} radio modules for \ac{IoT}, satellite based \ac{IoT} networks have become a feasible choice for mass deployments. Performance analysis of a novel architecture for \ac{IoT} nodes which do not have much computational resources is done. It is found that both \ac{SC} and \ac{MRC} schemes have same diversity gain bur different coding gain. This makes \ac{MRC} scheme perform better than \ac{SC} scheme. The proposed architecture suits present era of burgeoning number of \ac{LEO} satellite constellations and makes clever use of all the available satellite resources. In future, we intend to study this architecture by incorporating aspects like optimum power allocation between \ac{IoT} node and satellite terminals and effect of interference form nearby satellites and other radio devices.

\appendices
\section{Derivation of \eqref{op_ss_closed}}
\label{AppA}

\begin{figure}[t!]
\centerline{\includegraphics[width=2.8in,height=2.5in]{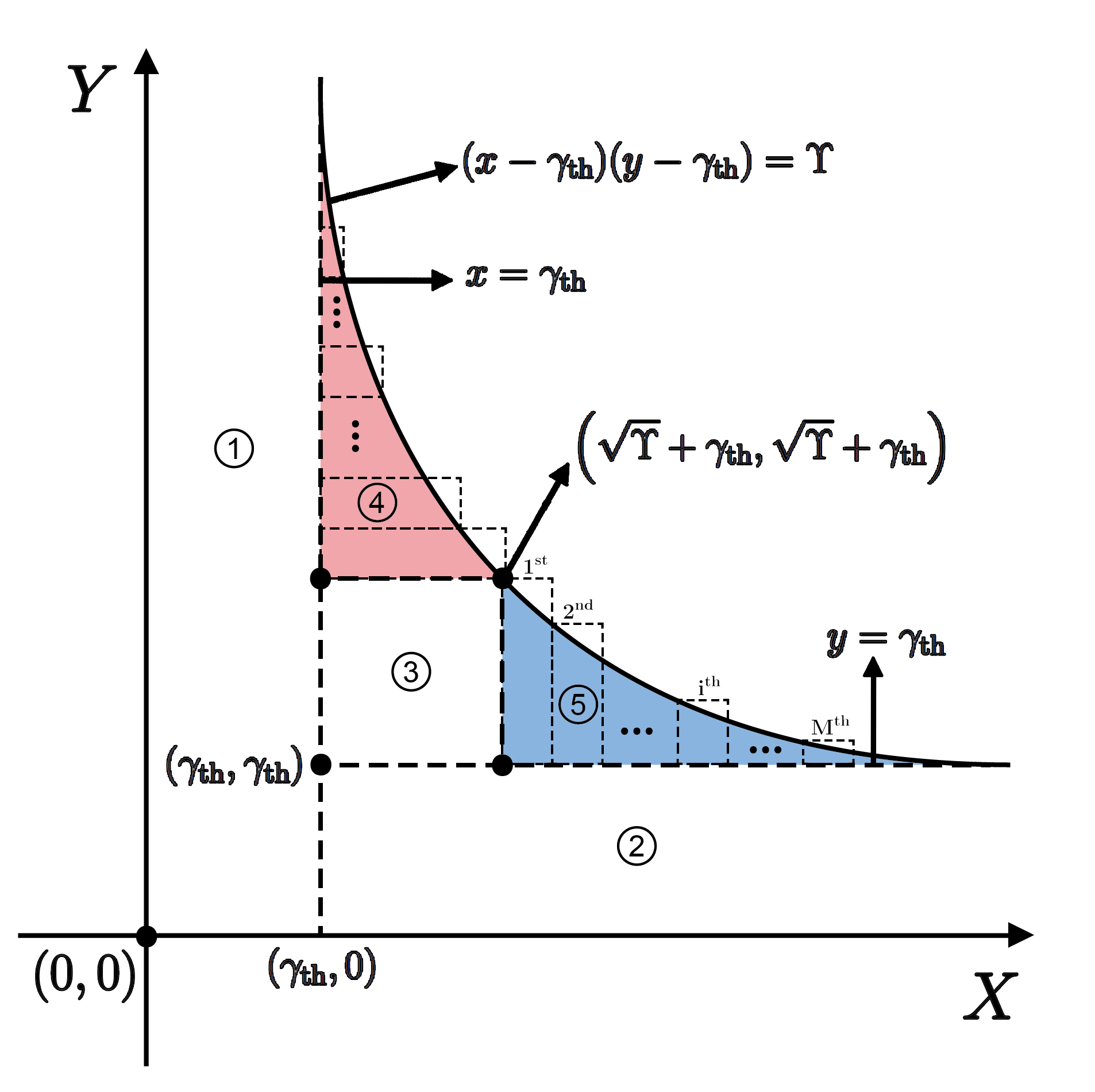}}
\vspace{-0.3cm}
\caption{Different regions of integration for calculating exact outage probability expression in scheme-1. Regions 1, 2 and 3 are calculated by direct integration, step staircase approximation is applied for only regions 4 and 5.}
\vspace{-0.5cm}
\label{step}
\end{figure}

To obtain the closed-form expression, we have to solve an equation of the form
\begin{equation}
P_{\text{out}} = \Pr\left[ (X - \gamma_{\text{th}}) (Y - \gamma_{\text{th}})\ \leq\ \Upsilon \right]
\label{step_stair}
\end{equation}
where $X, Y$ and $\Upsilon$ are defined from \eqref{op_ss1} as $X = \Lambda_{s_kg}$, $Y = \Lambda_{ns_k}$ and $\Upsilon = \gamma_{\text{th}}^2 + \gamma_{\text{th}}$, respectively. To solve the above expression, we can approximate the integral in five regions as shown in Fig. \ref{step}. For regions 4 and 5, following the process given in \cite{step}, we divide the integral region into $M$ vertical blocks. We divide region 4 into $M$ blocks from $Y=\sqrt{\Upsilon} + \gamma_{\text{th}}$ to $Y=\sqrt{\Upsilon} + \gamma_{\text{th}} + (L\times\gamma_{\text{th}})$, where $L$ is the depth of integration. Similarly we divide region 5 into $M$ blocks from $X=\sqrt{\Upsilon} + \gamma_{\text{th}}$ to $X=\sqrt{\Upsilon} + \gamma_{\text{th}} + (L\times\gamma_{\text{th}})$.  It is also important to note that, since $X$ and $Y$ are independent random variables i.e., $f_{X,Y}(x,y) = f_X(x)f_Y(y)$,  the integral values for regions $R_1$ to $R_3$, and $i$th block of $R_4$ and $R_5$ can be evaluated as
\begin{align}
\!\!R_1 = &\int_{y=0}^{\infty}\int_{x=0}^{\gamma_{\text{th}}}f_{X,Y}(x,y)dxdy = F_{X}(\gamma_{\text{th}}).
\label{r1} \\
\!\!\!R_2 = &\int_{y=0}^{\gamma_{\text{th}}}\int_{x=\gamma_{\text{th}}}^{\infty}\!\!f_{X,Y}(x,y)dxdy {=} F_{Y}(\gamma_{\text{th}})\left[1 {-}   F_{X}(\gamma_{\text{th}}) \right]\!.\!\!\!
\label{r2} \\
\!\!R_3 = &\int_{y=\gamma_{\text{th}}}^{\sqrt{\Upsilon}+\gamma_{\text{th}}}\int_{x=\gamma_{\text{th}}}^{\sqrt{\Upsilon}+\gamma_{\text{th}}}f_{X,Y}(x,y)\ dxdy\nonumber\\
 {=} &\left[\!F_{X}(\sqrt{\Upsilon} {+} \gamma_{\text{th}}) {-} F_{X}(\gamma_{\text{th}})\right]\!\! \left[\!F_{Y}(\sqrt{\Upsilon} {+} \gamma_{\text{th}}) {-} F_{Y}(\gamma_{\text{th}})\right]\!\!.\!\!\!
\label{r3}\\
\!\!R_{4}^{i} = &\int_{y=\sqrt{\Upsilon} + \gamma_{\text{th}} + \frac{i-1}{M}L}^{\sqrt{\Upsilon} + \gamma_{\text{th}} + \frac{i}{M}L}\int_{x=\gamma_{\text{th}}}^{\frac{\Upsilon}{\sqrt{\Upsilon} + \frac{i-1}{M}L} + \gamma_{\text{th}}}f_{X,Y}(x,y)\ dxdy\nonumber\\
 = &\left[F_X\!\left(\!\frac{\Upsilon}{\sqrt{\Upsilon} {+} \frac{i{-}1}{M}L} {+} \gamma_{\text{th}}\right) {-} F_X(\gamma_{\text{th}})\right]\nonumber\\
& {\times}\! \left[F_Y\!\left(\!\!\sqrt{\Upsilon} {+} \gamma_{\text{th}} {+} \frac{iL}{M}\right) {-} F_Y\!\left(\!\!\sqrt{\Upsilon} {+} \gamma_{\text{th}} {+} \frac{(i{-}1)L}{M}\!\right)\!\right]\!.\!\!\!
\label{r4}\\
\!\!R_{5}^i = &\int_{y=\gamma_{\text{th}}}^{\frac{\Upsilon}{\sqrt{\Upsilon} + \frac{i-1}{M}L} + \gamma_{\text{th}}} \int_{x=\sqrt{\Upsilon} + \gamma_{\text{th}} + \frac{i-1}{M}L}^{\sqrt{\Upsilon} + \gamma_{\text{th}} + \frac{i}{M}L} f_{X,Y}(x,y)\ dxdy\nonumber\\
 = &\left[F_Y\left(\!\frac{\Upsilon}{\sqrt{\Upsilon} {+} \frac{i{-}1}{M}L} {+} \gamma_{\text{th}}\right) {-} F_Y(\gamma_{\text{th}})\right]\nonumber\\
& {\times}\! \left[F_X\!\!\left(\!\sqrt{\Upsilon} {+} \gamma_{\text{th}} {+} \frac{iL}{M}\right) {-} F_X\!\left(\!\!\sqrt{\Upsilon} {+} \gamma_{\text{th}} {+} \frac{(i{-}1)L}{M}\!\right)\!\right]\!.\!\!\!
\label{r5}
\end{align}
Finally, the final closed-form expression for outage probability is given by
\begin{equation}
P_{\text{out}} = R_1 + R_2 + R_3 + \sum_{i=1}^{M}R_{4}^i + \sum_{i=1}^{M}R_{5}^i.
\label{step_final}
\end{equation}

\section{CDF of sum of \ac{SR} random variables}
\label{AppB}
The CDF of $\Delta_{i}, i\in\{sg,ns\}$ which is the sum of $K$ \ac{SR} random variables, is given by \cite{pdf_mrc}
\begin{equation}
\begin{split}
F_{\Delta_{i}}(x) =\ & \alpha_i^K \sum_{l=0}^{c}{\binom{c}{l}}\beta_i^{c-l}(\mathcal{G}(x,l,d,\eta)\\
&+ \epsilon\delta_i \mathcal{G}(x,l,d+1,\eta)),
\end{split}
\label{cdf_mrc}
\end{equation}
where $c = (d - K)^{+}, \epsilon = m_i K - d, d = \max\{K, \left \lfloor{m_iK}\right \rfloor \}$. $\mathcal{G}(x,l,d,\eta)$ is given as
\begin{equation}
\begin{split}
\mathcal{G}(x,l,d,\eta) =\ &\frac{(\beta_i - \delta_i)^{\frac{l-d-1}{2}}}{\eta^{\frac{d-l-1}{2}}\Gamma(d-l+1)} x^{\frac{d-l-1}{2}} e^{-\frac{\beta_i - \delta_i}{2\eta}x}\\
& \times M_{\frac{d+l-1}{2}, \frac{d-l}{2}}\left(\frac{\beta_i - \delta_i}{\eta}x\right),
\end{split}
\label{scriptf}
\end{equation}
where $M_{\mu,\nu}(\cdot)$ represents the Whittaker function\cite{formula}.

\bibliographystyle{IEEEbib}
\bibliography{ref}
\end{document}